\documentclass[a4paper,fleqn,usenatbib]{mnras}

\usepackage{newtxtext,newtxmath}
\usepackage{pdflscape}

\usepackage[T1]{fontenc}
\usepackage{ae,aecompl}
\usepackage{booktabs,threeparttable}

\usepackage{graphicx}	
\usepackage{amsmath}	
\usepackage{amssymb}	

\newcommand{\bv}{\begin{verbatim}} 
\newcommand{\be}{\begin{enumerate}}
\newcommand{\ev}{\end{verbatim}} 
\newcommand{\ee}{\end{enumerate}}
\newcommand{\vpfeiler}[1]{\rule[0em]{0em}{#1}}
\newcommand{\pC}{$m_{\rm C}$}
\newcommand{\pL}{m$_{\rm L}$}
\newcommand{\1}{{\it (i)}}
\newcommand{\2}{{\it (ii)}}
\newcommand{\3}{{\it (iii)}}
\newcommand{\4}{{\it (iv)}}

\newcommand{\chisq}{$\chi^2_{r}$} 
 \title[POLAMI II: Widespread circular polarisation]
        {POLAMI: Polarimetric Monitoring of Active Galactic Nuclei at Millimetre Wavelengths\\
        {\LARGE II. Widespread circular polarisation}}                          

\author[C. Thum et al.]{
            Clemens Thum$^{1}\thanks{E-mail: thum@iram.es (CT)}$,
            Iv\'an Agudo$^{2}$\thanks{E-mail: iagudo@iaa.es (IA)},
            Sol N. Molina$^{2}$,
            Carolina Casadio$^{3,2}$,
             \newauthor
            Jos\'e Luis G\'omez$^2$, 
            David Morris$^4$, 
            Venkatessh Ramakrishnan$^{5,6}$
            and Albrecht Sievers$^{1}$
\\
\\
$^{1}$Instituto de Radio Astronom\'ia Millim\'etrica, 
                Avenida Divina Pastora, 7, Local 20, E--18012 Granada, Spain\\
$^{2}$Instituto de Astrof\'{\i}sica de Andaluc\'{\i}a (CSIC),  
                 Apartado 3004, E--18080 Granada, Spain\\
$^{3}$Max-Planck-Institut f\"ur Radioastronomie, 
                Auf dem H\"ugel, 69, D--53121, Bonn, Germany\\
$^{4}$Institut de Radioastronimie Millim\'etrique,
                300 rue de la Piscine, Domaine Universitaire 
                38406 Saint Martin d'H\`eres, France,\\
$^{5}$Aalto University Mets\"ahovi Radio Observatory, 
                Mets\"ahovintie 114, 02540, Kylm\"al\"a, Finland\\
$^{6}$Universidad de Concepci\'on,  Departamento de Astronom\'ia, 
               Casilla 160-C, Concepci\'on, Chile
               }

\date{Accepted XXX. Received YYY; in original form ZZZ}

\pubyear{2017}

\begin{document}
\label{firstpage}
\pagerange{\pageref{firstpage}--\pageref{lastpage}}
\maketitle

\begin{abstract}
We analyse the circular polarisation data accumulated in the first 7 years of the POLAMI project introduced in an accompanying paper 
(Agudo et al.). 
In the 3mm wavelength band, we acquired more than 2600 observations, and all but one of our 37 sample sources were detected, most of them several times. 
For most sources, the observed distribution of the degree of circular polarisation is broader than that of unpolarised calibrators,
indicating that weak ($\lesssim0.5$\%) circular polarisation is present most of the time.  
Our detection rate and the maximum degree of polarisation found, 2.0\%, are comparable to previous surveys, all made at much longer wavelengths. We argue that the process generating circular polarisation must not be strongly wavelength dependent, and we propose that
the widespread presence of circular polarisation in our short wavelength sample dominated by blazars is mostly due to Faraday conversion of the linearly polarised synchrotron radiation in the helical magnetic field of the jet.
Circular polarisation is variable, most notably on time scales comparable to or shorter than our median sampling interval of $\lesssim1$ month. Longer time scales of about one year are occasionally detected, but severely limited by the weakness of the signal.
At variance with some longer wavelength investigations we find that the sign of circular polarisation changes in most sources, while only 7 sources, including 3 already known,  have a strong preference for one sign. 
The degrees of circular and linear polarisation do not show any systematic correlation. We do find however one particular event where the two polarisation degrees vary in synchronism during a time span of 0.9 years.
The paper also describes a novel method for calibrating the sign of circular polarisation observations.
\end{abstract}




\section{Introduction}
\label{Intr}

Astrophysical jets from active galactic nuclei (AGN) are powerful emitters of synchrotron radiation which can be strongly linearly polarised.
Its weak intrinsic circular polarisation (CP) has in the optically thin regime  a $\nu^{-0.5}$ spectrum 
\citep{Ginzburg1965,leggWestfold,jonesOdell,jones88}
which tends to make it negligible at short millimetre wavelengths. It was however realized early on \citep{sazonov}
that CP could also be generated by Faraday conversion (FC). This mechanism can convert the strong linear polarisation to circular in two ways which exploit the birefringence of the magnetised plasma of the jet. Both ways require a departure
from the original angle between the magnetic field direction and the polarisation which is $0^\circ$ or $90^\circ$  for  high or low optical depths, respectively. Either \1 the polarisation angle is changed by Faraday rotation (FR) in the emitting medium (FR--driven FC; \citet{melrose71}) or \2 the direction of the magnetic field {\bf B} rotates along the propagation direction (BR--driven FC; \citet{kennettMelrose}). The rotation of the {\bf B}--direction can be due to a stochastic field component \citep{marscherTurbulence}
or to  a changing large scale field organisation like in helical fields \citep{gabuzda2008}.  These topics are discussed in detail by \citet{WH2003}, 
together with other open issues like the composition of the jet, its density structure, and the wavelength dependence of the resulting CP. 
  
Studies of circular polarisation have therefore great potential in contributing to the understanding of the still poorly known fundamental properties of AGN jets. Due to the weakness of circular polarisation \citep{weilerDePater1983,saikiaSalter,homanLister2006}, however, 
sensitive and well calibrated instrumentation at big telescopes is needed,
and the number of investigations are not many. Single dish surveys have been made with the Parkes telescope \citep{komesaroff1984}, the University of Michigan radio telescope \citep{aller2003}, the Effelsberg 100m telescope \citep{F-GAMMA,Myserlis}, the Arecibo telescope (R. Taylor, private communication),
and the IRAM 30m telescope \citep{survey1,survey2}. 
Interferometric observations were made with the VLBA, in particular the MOJAVE survey \citep{homanLister2006}, and with the VLA \citep{bowerLLAGN2002} and the ATCA \citep{rayner2000,osullivan2013}. For a more complete summary refer to \citet{Homan2009}.
Another important factor is the opacity of the jet plasma. Usually, observations need to be made at short, often millimetre, wavelengths where the opacity is sufficiently low for the jet core and emerging new components to be seen \citep{Lobanov1998}.     
The often rapid temporal variations of the jet emission further complicate matters. These variations originate in energetic events near the black hole 
and then travel downstream of the jet.
In order to resolve these highly dynamic events, adequately fast time resolution is needed.

Our program POLAMI (Polarimetric Monitoring of AGN at Millimetre wavelengths) project addresses all these issues. We exploit the high sensitivity available with the IRAM 30m telescope; we observe at 3 and 1.3mm, the shortest wavelength of any CP monitoring program so far; and we observe a large sample of 37 sources
 once every month.  A further distinguishing property of POLAMI is its capability of simultaneously measuring all four Stokes parameters 
at both wavebands in parallel, a very useful feature when interpreting CP events.  Observing procedure, data reduction, and sample characteristics are described in \citet{paperI}, 
henceforth Paper~I. Here we analyse the circular polarisation data obtained during the first seven years of the monitoring
campaign. The total density and linear polarisation data, and their variability properties
are analysed  separately 
\citep{paperIII}.  Detailed studies of specific sources and discussions of
other statistical aspects of  our sample, like correlations with optical, $\gamma$--ray, and 7mm VLBI data will be presented in future publications.


\section{General properties of the data set}
\label{s:data}

The circular polarisation data used here are taken from the database of the
POLAMI (Polarimetric Monitoring of AGN at Millimeter Wavelengths)
program described in the accompanying 
Paper~I. 
The astrophysical properties of the sources, notably their positions, optical classification, spectral energy distribution, and redshift are described in 
Paper~I\footnote{
We keep the linear polarisation calibrator 1328+307 in the sample, since its circular polarisation properties are largely unexplored.
}. 
The sample  of the more frequently observed sources comprises 37 AGN, mostly quasars (22) and BL\,Lac objects (11), and a few radio galaxies (4). All of these sources have at least 25 valid observations in the 3mm band. The average number of observations is 61 per source.

\begin{figure}
\centering
\includegraphics[width=0.48\textwidth]{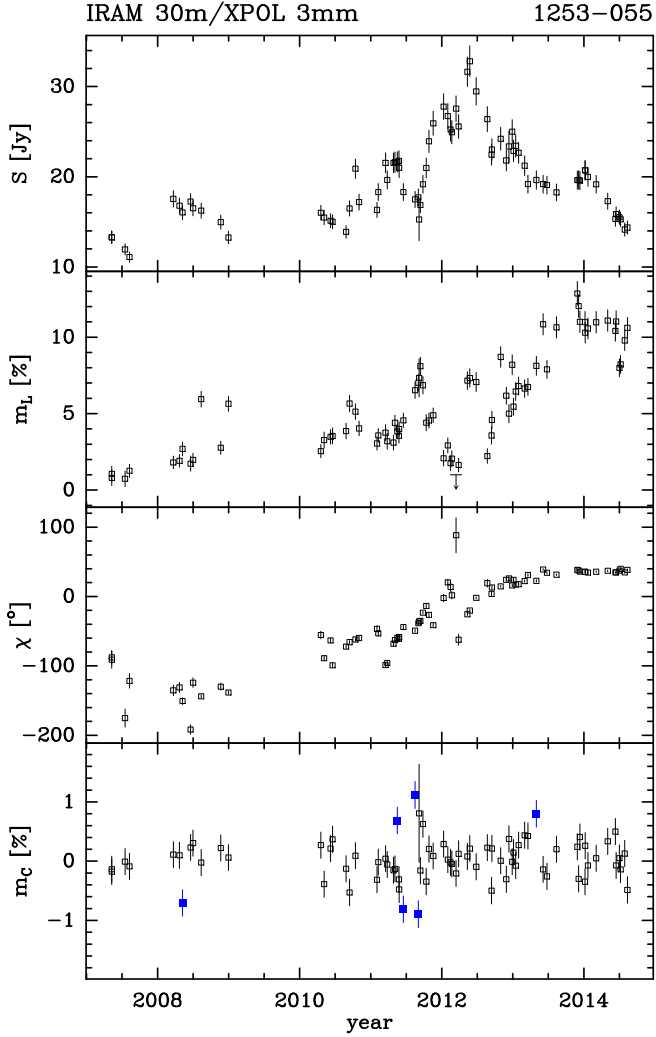}
   \caption{
     The quasar 1253$-$055 at 3mm ($z = 0.538$). The 4 frames
     show, from top to bottom, the total flux density (Stokes I), the degree of
     linear polarisation, \pL;  the polarisation angle, $\chi$, running
     from  north to east in the equatorial system;
     and the degree of circular polarisation, \pC, which is detected at 6
     epochs at S/N $\geq3$ (blue filled squares). 
     Data and figure are from the POLAMI database 
(Paper~I). 
           }
      \label{f:DBexample}
\end{figure}

The observations were made with the IRAM 30m Telescope at 3mm and 1.3mm
wavelengths using the procedure XPOL \citep{XPOL} which acquires
simultaneously the four Stokes parameters. 
The angular resolution at these wavelengths is 28'' and 12'', respectively. For
more technical details we refer to 
Paper~I. Monitoring started on
14--October--2006 at 3mm wavelength using the Observatory's ABCD receivers,  and from 07--Dec--2009 on with the receiver EMIR \citep{EMIR} 
which permitted simultaneous recording of 3 and 1.3mm bands.
Observations were made at irregular intervals, typically of 2 -- 4 weeks.
Here we discuss the circular polarisation data obtained up to August 2014. Figure~\ref{f:DBexample} shows an example of the full polarisation data
obtained for one of the monitored sources at 3mm wavelength.


\subsection{Precision of measurements}
\label{ss:noise}

Calibration of the observed Stokes Parameters and the correction of $Q, U,$   and $V$ for instrumental polarisation (IP) were described in 
Paper~I. 
Here we apply an additional correction which captures a slow drift of the instrumental $V$ due to the coarse time steps of the principal IP correction.
The amplitude of the additional correction, sampled at 3--month steps, is small ($\lesssim0.5$\%). It is recognizable only in Stokes $V$ which is the Stokes parameter best determined with XPOL \citep[see][]{XPOL}.

We use observations of the unpolarised  sources Mars and Uranus for the measurement of the slow drift. Several independent observations are usually made in each 3--month interval. In a second step we use the mean of the brighter half of our AGN sample for an alternative derivation of the slow drift. Although polarised individually, the mean CP  of an AGN sample should be unpolarised if a large enough number of observations of different sources is averaged, which is the case here.  Both methods agree within their errors with the AGN giving a more precise result because of the much larger number of observations.  
In the 3mm band, the overall rms scatter of $V$ is reduced for these calibrators from 0.34\% to 0.22\% after the subtraction of the slow drift. The reduction at 1mm is however negligible
, due to the higher statistical errors at this wavelength.  
The final degree of circular polarisation \pC\  is then obtained from the IP--corrected observation of Stokes $V$ by subtracting this slow drift only at 3mm. 
The resulting time series of circular polarisation is shown in 
Fig.~\ref{f:time1} for all sample sources  at  both wavebands. 
This Figure also includes the calibrators Mars and Uranus.


\begin{figure*}
\centering
\includegraphics[width=1.0\textwidth]{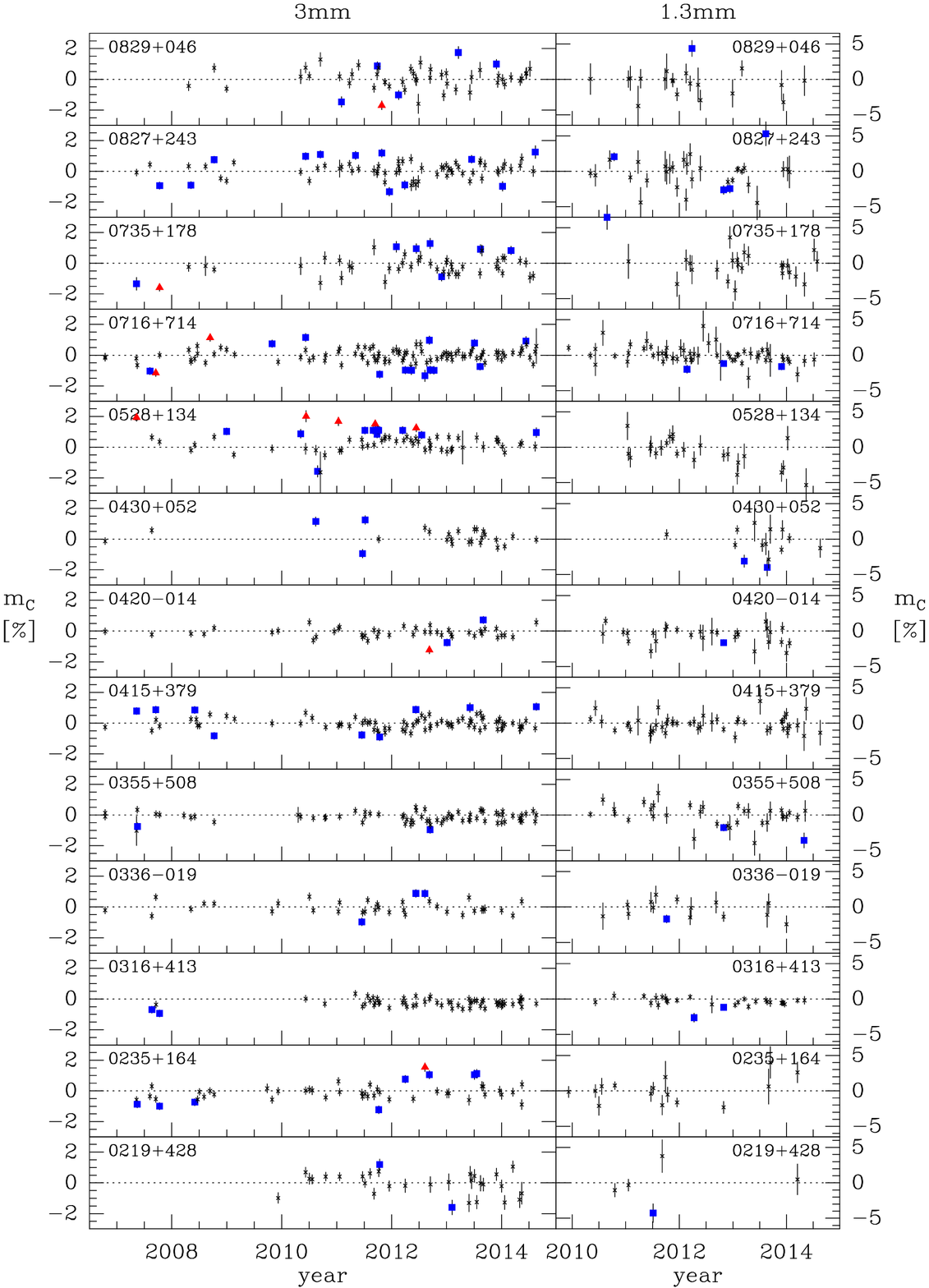}
   \caption{Time sequences of 3mm (left) and 1mm (right) observations of 
            all 37 sources in our sample.
            The degree of circular polarisation \pC\ is shown in percent with
            its $1\sigma$ error bar. Detections are shown as filled symbols 
            (blue squares for S/N $\geq3$ or red triangles for S/N $\geq5$).
           }
   \label{f:time1}
\end{figure*}

\setcounter{figure}{1}
\begin{figure*}
\centering
\includegraphics[width=1.0\textwidth]{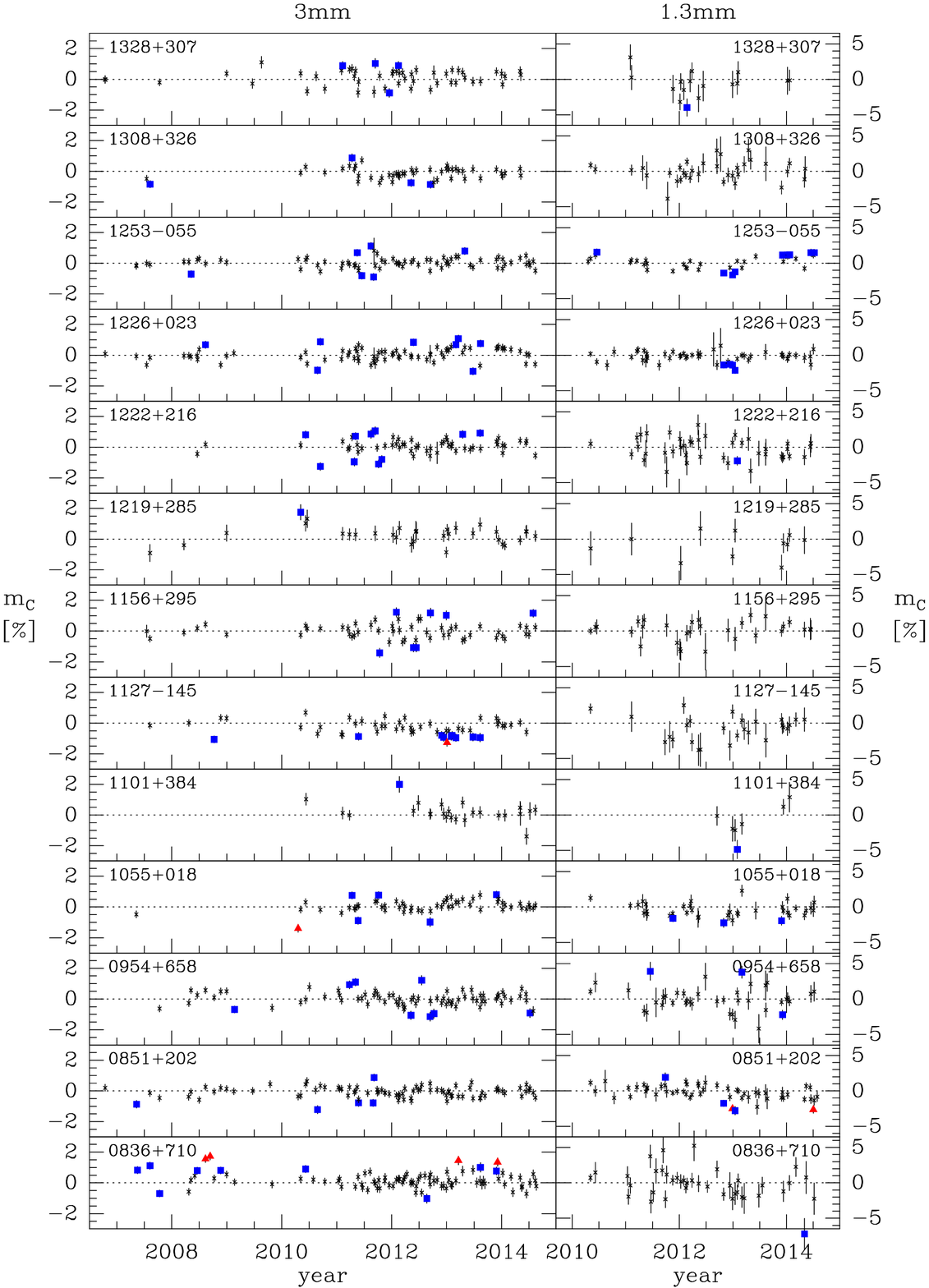}
   \caption{Continued.}
\end{figure*}

\setcounter{figure}{1}
\begin{figure*}
\centering
\includegraphics[width=1.0\textwidth]{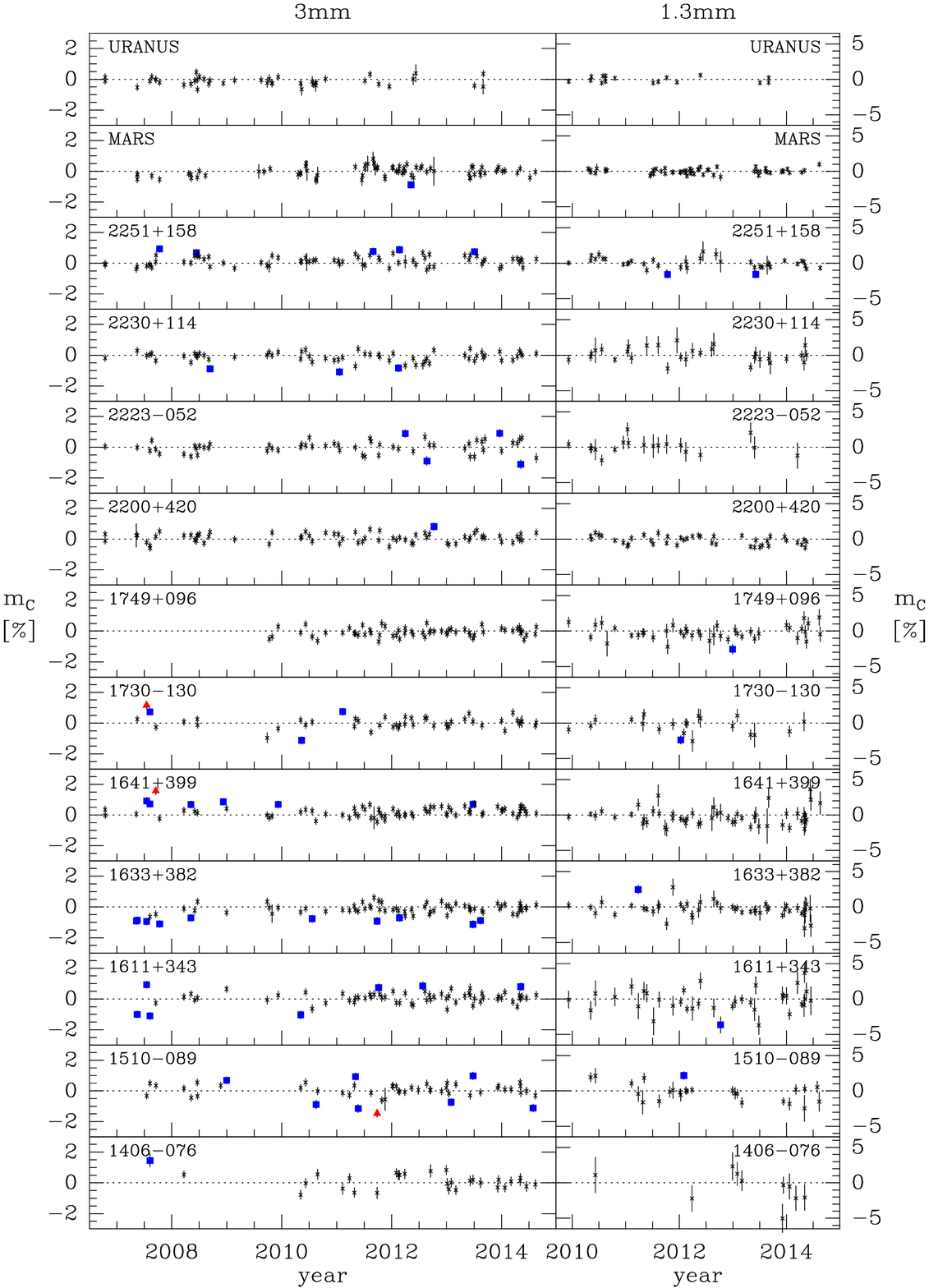}
   \caption{Continued.}
\end{figure*}
%

\begin{figure}
\centering
\includegraphics[width=0.47\textwidth]{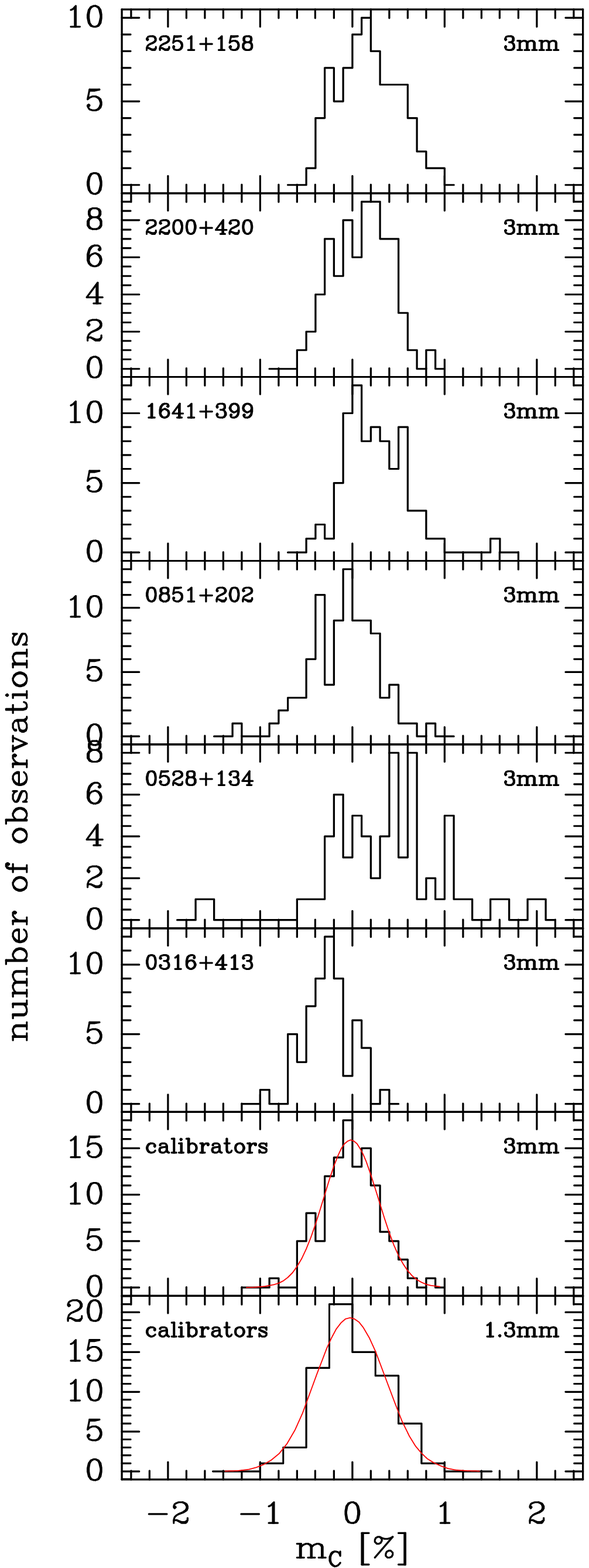}
   \caption{Histograms of the distribution of \pC\ measurements at
            3mm wavelength for five selected AGN and the calibrators
            (Mars and Uranus combined) at 3 and 1.3 mm. 
            Red curves are Gaussian fits.
            Histograms of the remaining sample sources can be found online.
           }
   \label{f:histograms}
\end{figure}

The errors assigned to individual CP observations contain the statistical
error of the Stokes $V$ and $I$ parameters and a systematic error estimated from the rms scatter of the calibrator observations described above, added in quadrature. Except for observations of very low total flux density, the systematic error is always dominant in 3mm observations. In the 1mm band, the systematic error is larger than the statistical error for observations where the total flux density is above 2.0 Jy. This is the case for 30\% of the observations.
Quantitatively, we take the systematic error from the 
histograms  in Fig.~\ref{f:histograms}, for which data of the unpolarised Mars and Uranus were combined.  The histograms  are fitted by Gaussians, and we derive standard deviations of $0.28\pm0.02$\,\% (3mm) and $0.37\pm0.03$\,\% (1.3mm).
For the 3mm band, using a typical value of the statistical noise (0.11\%), this results in a  mean $3\sigma$ detection threshold of 0.90\,\%. 
The higher systematic error in the 1mm band together with the much higher  thermal noise due to the 3--5 times higher system temperature result in a 
mean $3\sigma$ detection threshold of 1.5\% for a 2 Jy source and higher for the weaker ones.

\begin{figure}
\centering
\includegraphics[width=0.48\textwidth]{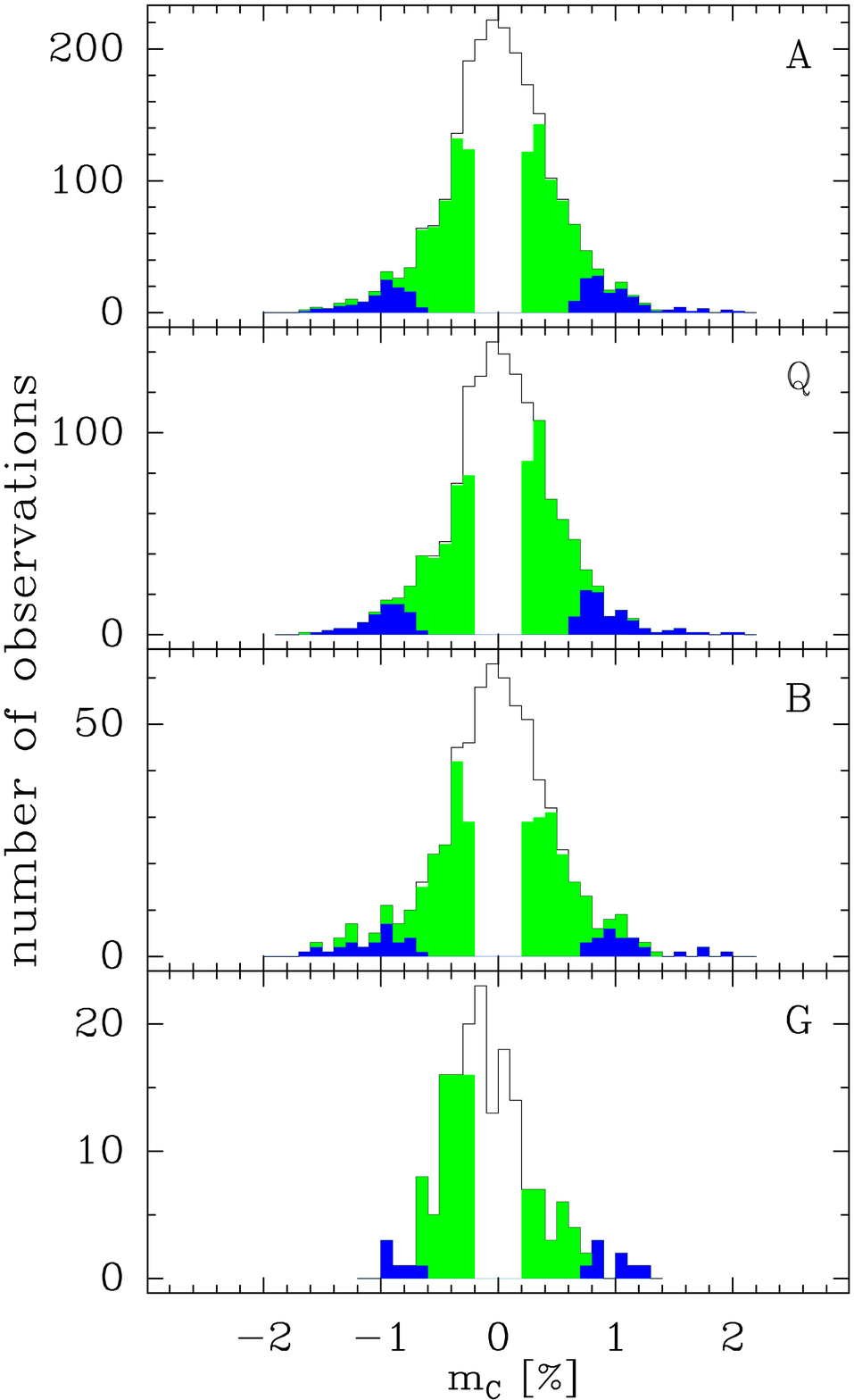}
   \caption{Histogram of  circular polarisation observations. 
     The uppermost frame, labelled A, shows all observations (2261)
     obtained in our monitoring program. Frames below display
     different AGN classes: quasars (Q), BL\,Lac objects (B), and
     radio galaxies (G). Colours code measurements of different     
     signal--to--noise ratios:
     blue ($\geq3$), green ($\geq1$). White space
     indicates observations at S/N < 1.  
           }
      \label{f:masterHisto}
\end{figure}

Inspection of the time series of the calibrators (Fig.~\ref{f:time1}) indicates that the applied systematic errors do not always lead to a reliable estimate of the total errors.
This is particularly the case for observations made during summer afternoons
when the atmosphere is often disturbed by anomalous refraction.
We have screened the database for observations taken under such conditions and rejected any unreliable observations. 
The bulk of the database is however unaffected by anomalous refraction and we consider a signal--to--noise ratio S/N = 3 a reliable statistical detection threshold. For individual detections we require S/N $\geq5$, as is common practice in massive data flows.

The global properties of our observations are further illustrated in Fig.~\ref{f:masterHisto} where the top frame labelled A shows the histogram of  the 3mm measurements of CP in all sample sources, irrespective of source type.  The coloured bins indicate the observations with signal--to--noise exceeding  1 (green) and 3 (blue), while the bell--shaped envelope refers to all 2261 observations. The envelope is very symmetric and well centred on zero. Apparently, no systematic bias is introduced in the data acquisition or reduction. The small asymmetry in the panel labelled ``G'' (radio galaxies) is exclusively due to 0316+413 (3C\,84), a source already known for its strongly biased \pC\ distribution (see Sect.~\ref{ss:signs} and 
Figs.~\ref{f:histograms} and \ref{f:time1}).
Quantitatively, the mean of all data is \pC\ = $+0.014\pm0.010$ \% and the rms scatter is $\sigma = 0.40\pm0.01$ \%.  There are 34 more measurements with positive \pC, corresponding to a negative/positive imbalance of only 0.75\%.  The rms scatter of these measurements
is significantly larger than that of the unpolarised calibrators  ($0.28\pm0.02$), as
derived above. Clearly, circular polarisation is widespread in our sample, even if our sensitivity allows firm individual detections only for the top 1\% of events (Tab.~\ref{t:summary} and Sect.~\ref{ss:rates}).

\subsection{Calibration of the sign of Stokes $V$}
As there is no celestial calibrator of circular polarisation available at short millimetre wavelengths, calibration of the sign of Stokes~$V$ is not straightforward. We employed a novel method, described in detail in Appendix \ref{s:crab}, using the known transport of the strongly linearly polarised 3mm signal from the Crab nebula through the telescope's Nasmyth optics. At a suitable location in the receiver cabin a quarter wave plate optimised for 3mm wavelength was placed such that its fast axis made angles of $\pm45^\circ$ with the known direction of linear polarisation of the Crab at two specific hour angles. 
The resulting left- or right-handed circular polarisation provides the calibration of the sign of Stokes~$V$. This calibration is independent of wavelength, as long as observations are made with identical setup of frequency downconversion.

We confirmed this sign calibration by making observations of the radio star MWC\,349. This strong source of millimetre recombination lines is known to have a magnetic field which introduces circular polarisation at the level of 1\% in its H30$\alpha$ line emission \citep{ThumMorris}. Within the uncertainties associated with the time variability of the signal \citep{Stellenbosch} the sign obtained agrees with previous observations. This strongly supports our present calibration, since the first observation by \citet{ThumMorris} used completely different equipment whose calibration of the Stokes~$V$ sign was however straightforward. MWC\,349 may thus qualify as a calibrator for the sign of circular polarisation at short mm, and possibly also at submm wavelengths.  

In an unrelated program (Agudo et al. 2017, in prep.) we monitored the Galactic Centre source Sgr\,A$^\ast$ with similar equipment as used for our POLAMI program. Circular polarisation was detected throughout this campaign, consistently of negative sign. This is in agreement with many earlier 
detections 
\citep{bowerCpSgrA1999,saultMacquart1999,bowerCpSgrA2002,munoz2012}, and further confirms the sign of our Stokes~$V$ calibration. 

\section{Results}

During the 7 years of our monitoring program we obtained more than 3300 observations in the 3mm band and  more than 2700 at 1mm. Tab.~\ref{t:summary} gives 
the corresponding numbers of {\it valid} observations which are lower due to occasional technical problems and observational errors, but mostly due to atmospheric instabilities or high water vapour. These affect the 1mm band more, explaining the much lower number of valid 1mm observations together with the lack of 1mm receiver during the first four years of the monitoring period.
In the 3mm band, more than half of the sources in the sample have at least one strong (S/N $\geq5$) detection, compared to only one at 1mm. Altogether, 31 strong detections were made at 3mm against only one at 1mm. The S/N $\geq3$ detections give a similar difference between the two bands.

This order of magnitude reduction of detections is fully accounted for by the higher 1\,mm detection threshold (factor 1.7) and the lower number of valid observations (factor 3.7), suggesting that the intrinsic frequency and properties of circular polarisation events may be quite similar at the two wavebands. The data in this table already demonstrates that circular polarisation at short mm wavelengths is widespread among blazars and  highly variable.  It is detectable at 3mm  wavelength with relative ease, but our sensitivity in the 1mm band is barely adequate.

\subsection{Detection rates}
\label{ss:rates}

\begin{table}
\caption{Summary of sources observed and detected in circular polarisation at each waveband.
  }
\label{t:summary}      
\centering  
\begin{tabular}{l c c} 
\hline             
number of                                & 3mm     & 1mm \vpfeiler{2.5ex}\\[0.5ex]\hline
sources in sample                        & 37      & 37   \vpfeiler{2.0ex}\\ 
observations                             & 2261    & 619    \\
sources detected at S/N $\geq5 / 3$      & 20 / 36 & 1 / 14   \vpfeiler{0.0ex}\\
individual detections at S/N  $\geq5 / 3$ & 31 / 267& 1 / 31   \\[0.5ex]\hline 
\end{tabular} 
\end{table}

From the numbers given in Tab.~\ref{t:summary} we calculate 
detection rates  of 11.8\%  and 5.0\% for the 3 and 1\,mm bands.
Detection rates reported in our previous  single epoch surveys are
6\% \citep{survey1} and 8\% \citep{survey2} in the 3mm band. 
These numbers are roughly comparable, especially when taking into account the lower sensitivity of the first survey. At 1mm, the only previous single epoch survey \citep{survey2} reports no detections. 

The single epoch MOJAVE survey \citep{homanLister2006} observing at 15 GHz with milliarcsecond resolution 
detected 17  out of their sample of 133 sources, resulting in a detection rate of 13\%,  nearly equal to our detection rate at 3mm. This agreement is rather surprising, since 
there are at least two factors which tend to put our 3mm survey at a
disadvantage: {\it (i)} their higher polarisation sensitivity (0.3\%  vs. 0.1\%) and  {\it (ii)} beam depolarisation of CP in our much larger beam.
The similarity of the detection rates may however be understood if 
the lower 3mm opacity acts to select the jet components with highest brightness temperature in much the same way as does the high angular resolution of the MOJAVE survey.  We are also driven to the conclusion  that the tendency for beam depolarisation as observed in 3C84 \citep{HW2004} cannot be strong in our AGN sample. In fact, 3C84 is so far the only source where jet components of different CP sign are observed to be simultaneously present.

\subsection{Source statistics}
\label{ss:histograms}                       

Tab.~\ref{t:bigTable} summarises the results derived from the time series (Fig.~\ref{f:time1}). For each source in the sample and both wavebands we give 
\1 the total number of observations with valid Stokes~$V$ data,
\2 the median \pC, \3 the standard deviation of \pC,
and \4 the number of detections above signal--to--noise S/N $\geq5$ and $\geq3$, respectively.  For sources which have been detected previously in circular polarisation, we add the reference to the respective publications.


\begin{table*}
\caption{List of sources observed and their circular
         polarization properties. 
        }
\label{t:bigTable}      
\centering  

\begin{tabular}{c c c c c r@{\,}c@{\,}l  c  c c c r@{\,}c@{\,}l l}  
\hline              
 & common & \multicolumn{6}{c}{3mm} & & \multicolumn{6}{c}{1mm} & previous
   \vpfeiler{2.5ex}\\
  \cline{3-8} \cline{10-15}
source\,$^{a}$ \vpfeiler{2.5ex} &name&N$_{\rm obs}$$^{b}$ & 
      \multicolumn{1}{c}{\,$m_{\rm C}\,^{c}$} &
      \multicolumn{1}{c}{\,$\sigma\,^{d}$} &\multicolumn{3}{c}{N$_{\rm det}$$^{e}$} & &
      N$_{\rm obs}$$^{b}$  & \multicolumn{1}{c}{\,$m_{\rm C}\,^{c}$} &     
      \multicolumn{1}{c}{\,$\sigma\,^{d}$} & 
      \multicolumn{3}{c}{N$_{\rm det}\,^{e}$} & detections \\[1ex]
      \hline \vpfeiler{2.5ex}
%
%

0219+428 & 3C\,66A &30 &0.01 & 0.73  & 0 &/&2
  &    &4 & -0.68  & 2.86 & 0&/&1
&\\
0235+164 & AO\,0235+164 &49 &-0.14 & 0.55  & 1 &/&8
  &    &14 & 0.19  & 1.69 & 0&/&0
&\\
0316+413 & 3C\,84 &59 &-0.25 & 0.25  & 0 &/&2
  &    &29 & -0.41  & 0.59 & 0&/&2
&1,5
\\
0336$-$019 & CTA\,26 &37 &-0.17 & 0.43  & 0 &/&3
  &    &15 & -0.10  & 1.16 & 0&/&1
&\\
0355+508 & NRAO\,150 &72 &-0.12 & 0.31  & 0 &/&2
  &    &35 & 0.09  & 1.53 & 0&/&2
&\\
0415+379 & 3C\,111 &88 &-0.05 & 0.42  & 0 &/&9
  &    &47 & -0.08  & 1.07 & 0&/&0
&\\
0420$-$014 & PKS\,0420-01 &46 &-0.13 & 0.35  & 1 &/&2
  &    &29 & -0.28  & 1.12 & 0&/&1
&\\
0430+052 & 3C\,120 &26 &0.05 & 0.50  & 0 &/&3
  &    &13 & -0.83  & 1.81 & 0&/&2
&\\
0528+134 & PKS\,0528+134 &67 &0.43 & 0.65  & 5 &/&11
  &    &22 & -0.89  & 1.52 & 0&/&0
&\\
0716+714 & S5\,0716+71 &108 &0.01 & 0.49  & 2 &/&13
  &    &62 & -0.05  & 1.19 & 0&/&3
&2,3
\\
0735+178 & OI\,158 &50 &-0.20 & 0.66  & 1 &/&7
  &    &21 & -0.18  & 1.62 & 0&/&0
&\\
0827+243 & OJ\,248 &70 &0.16 & 0.57  & 0 &/&12
  &    &33 & -0.03  & 2.24 & 0&/&5
&1
\\
0829+046 & OJ\,049 &51 &0.13 & 0.71  & 1 &/&5
  &    &19 & -0.11  & 1.82 & 0&/&1
&\\
0836+710 & 4C\,71.07 &91 &0.14 & 0.51  & 4 &/&9
  &    &39 & -0.02  & 2.14 & 0&/&1
&1,3
\\
0851+202 & OJ\,287 &89 &-0.05 & 0.36  & 0 &/&5
  &    &49 & -0.09  & 0.90 & 1&/&1
&5,6
\\
0954+658 & S4\,0954+65 &78 &0.02 & 0.47  & 0 &/&8
  &    &36 & -0.10  & 1.51 & 0&/&3
&2
\\
1055+018 & 4C &59 &0.04 & 0.41  & 1 &/&5
  &    &38 & -0.58  & 0.96 & 0&/&3
&3,6
\\
1101+384 & Mrk\,421 &25 &0.14 & 0.58  & 0 &/&1
  &    &7 & -1.27  & 2.23 & 0&/&1
&\\
1127$-$145 & PKS\,1127-14 &51 &-0.28 & 0.45  & 1 &/&9
  &    &27 & -0.61  & 1.67 & 0&/&0
&\\
1156+295 & 4C\,29.45 &58 &-0.01 & 0.55  & 0 &/&7
  &    &30 & 0.26  & 1.36 & 0&/&0
&3,6
\\
1219+285 & W\,Comae &32 &0.35 & 0.56  & 0 &/&1
  &    &11 & -0.58  & 1.71 & 0&/&0
&\\
1222+216 & 4C\,+21.35 &60 &0.13 & 0.48  & 0 &/&10
  &    &37 & -0.26  & 1.49 & 0&/&1
&\\
1226+023 & 3C\,273 &81 &0.06 & 0.42  & 0 &/&8
  &    &47 & -0.16  & 0.71 & 0&/&4
&3,4,5
\\
1253$-$055 & 3C\,279 &76 &0.03 & 0.35  & 0 &/&6
  &    &34 & 0.32  & 0.77 & 0&/&6
&3,4,5,6
\\
1308+326 & OP\,313 &51 &-0.15 & 0.38  & 0 &/&4
  &    &33 & -0.04  & 1.40 & 0&/&0
&\\
1328+307 & 3C\,286 &53 &0.23 & 0.50  & 0 &/&4
  &    &16 & -0.47  & 1.67 & 0&/&1
&2
\\
1406$-$076 & PKS\,1406-076 &30 &0.11 & 0.50  & 0 &/&1
  &    &8 & -0.02  & 2.24 & 0&/&0
&\\
1510$-$089 & PKS\,1510-08 &52 &0.06 & 0.49  & 1 &/&7
  &    &26 & -0.08  & 1.23 & 0&/&1
&6
\\
1611+343 & DA\,406 &68 &0.10 & 0.41  & 0 &/&7
  &    &36 & -0.18  & 1.60 & 0&/&1
&\\
1633+382 & 4C\,38.41 &76 &-0.17 & 0.37  & 0 &/&10
  &    &46 & -0.29  & 0.94 & 0&/&1
&3,6,7
\\
1641+399 & 3C\,345 &80 &0.20 & 0.33  & 1 &/&6
  &    &50 & -0.29  & 1.21 & 0&/&0
&5,7
\\
1730$-$130 & NRAO\,530 &48 &0.05 & 0.40  & 1 &/&3
  &    &21 & -0.14  & 1.07 & 0&/&1
&\\
1749+096 & 4C\,+09.57 &60 &-0.09 & 0.27  & 0 &/&0
  &    &43 & -0.26  & 0.97 & 0&/&1
&6
\\
2200+420 & BL\,Lacertae &70 &0.12 & 0.29  & 0 &/&1
  &    &42 & -0.16  & 0.60 & 0&/&0
&1
\\
2223$-$052 & 3C\,446 &59 &0.02 & 0.42  & 0 &/&4
  &    &19 & 0.15  & 0.98 & 0&/&0
&6
\\
2230+114 & CTA\,102 &64 &-0.06 & 0.32  & 0 &/&3
  &    &29 & -0.03  & 0.94 & 0&/&0
&6,7
\\
2251+158 & 3C\,454.3 &77 &0.16 & 0.32  & 0 &/&5
  &    &38 & -0.04  & 0.74 & 0&/&2
&1
\\    
\hline                                        
\end{tabular} 
\begin{tablenotes}
\item References of previous detections: (1)~\citet{survey1}; (2)~\citet{survey2}; (3)~\citet{homanLister2006};
(4)~\citet{weilerDePater1983}; (5)~\citet{aller2003}; (6)~\citet{Vitrishchak2008}; (7)~\citet{gabuzda2008}\\
\item Explanation of columns:
$^{a}$ IAU B1950 source name. J2000 positions are given in paper~I together with other source properties.
$^{b}$ number of observations at the respective wavelength.
$^{c}$ median value of circular polarization, in percent, over the monitoring period.
$^{d}$ rms of individual measurements of \pC\ over the monitoring period.
$^{e}$ number of detections (5$\sigma$/3$\sigma$) of circular polarization
(We take $3\sigma$ detections to have S/N between 3 and 5), while $5\sigma$ detections have S/N $\geq5$).     
\end{tablenotes}
\end{table*}


In the 3mm band, all  but one sources of our sample (1749+096) were detected, many of them several times. 
The record holder is the quasar 0528+134 with 16 detections in the 3mm band,
5 of which have S/N $\geq5$. 
The source with the most (109) observations, 0716+714, was detected 15 times. 
Detections in the 1mm band are an order of magnitude less numerous. 

Inspection of Tab.~\ref{t:bigTable} shows that for many sources 
the median of the \pC\ observations is offset from zero. Most sources have 
3mm \pC\ distributions whose width is significantly broader than that of the unpolarised calibrators. We emphasize that the larger width is not due to higher noise of the AGN observations. As stated in sect.~\ref{ss:noise}, the noise of all sources is dominated by the systematic uncertainties as derived from the calibrators, and  it is therefore very similar for all sources.
For better visualisation of these properties we built
histograms of the \pC\ distribution for all sources with more than 50
observations. Histograms with less observations are too noisy. At 3mm, 28 sources satisfy this criterion, but unfortunately, none at 1mm.

Fig.~\ref{f:histograms} shows these histograms for a few selected AGN together with the calibrators (Mars and Uranus combined). The \pC\ distributions of the AGN are clearly different from that of the calibrators. 
The sources 0316+413 and 1641+399 have \pC\ distributions not or barely broader than the calibrators, but offset from zero in opposite directions. Half a dozen more sources have similar offsets, positive and negative directions equally frequent.  No instrumental effect can produce such systematic offsets, since each histogram is assembled from observations randomly distributed over the 7 year monitoring period. However complex the intrinsic variability of blazars may be in general, there exists in a minority of sources a mechanism capable of maintaining  a CP bias over 7 years.  Other histograms, like those for  0851+202 or 2200+420, appear to be more nearly flat--topped rather than Gaussian. Such distributions may arise in sources where \pC\ moves randomly and rapidly between a positive and a negative limit. In some histograms there is weak evidence for a central dip near zero \pC\  as in 2251+158 as well as in $0336-019$, 0355+508, and 1101+384. Some other histograms, like in 0528+134, are quite irregular and not centrally peaked anymore. 

We quantified the difference between the targets and the calibrator by calculating the reduced \chisq\ of the distribution of all 28 sources using the  Gaussian of the calibrators as the parent distribution of an unpolarised source. The median value of these sources is \chisq\ =2.55, corresponding to a confidence level of better than 99.5\% that these histograms are different from that of an unpolarised source. Even the source with the smallest \chisq\ (1.50 for 2230+114 which has 64 observations, three of which have \pC $>3\sigma$) is different from the unpolarised
parent distribution at 95\% confidence. Clearly, all  well observed 3mm sources have \pC\ distributions indicative of significant circular polarisation.  This is also the case for our sample in general as shown by the histogram of all sample sources combined (Fig.~\ref{f:masterHisto}, top) and discussed in Sect.~\ref{ss:noise}.

\subsection{Maximum degree of circular polarisation} 
\label{ss:weak} 
The maximum |\pC| reliably detected at S/N$\geq5$ in our sample is 2.0\% at 3\,mm (Fig.~\ref{f:masterHisto}) and 2.6\% at 1\,mm.
CP stronger than 2.0\% was detected only once in each band, in 0528+134 (3mm) and in 0851+202 (1\,mm).
Comparing with previous CP surveys at lower frequencies, we find that our
|\pC|$_{max}$ is significantly higher than most maxima found before.
 An early compilation of CP observations by
\citet{weilerDePater1983} lists only two sources in their top |\pC| bin
between 0.4 to 0.5\%.  \citet{aller2003}
using the University of Michigan radio telescope at 4.8 and 8 GHz
monitored 16 blazars during many years. They also find a low maximum |\pC| =
0.9\%.  
The MOJAVE first epoch survey \citep{homanLister2006} made with the
VLBA at 15 GHz finds a maximum |\pC| of only 0.7\%. 
Using the same instrument and observing simultaneously at 15, 22, and 43 GHz
(7 mm) \citet{Vitrishchak2008} found \pC\ highest at their highest frequency.
 In a statistical sense, it appears that |\pC| tends
to increase  with frequency, with our survey marking the present
maxima of 2.0\% and 2.6\% at 3 and 1\,mm wavelengths, respectively.  

Specific low frequency (cm wavelength) studies of a few objects have however also turned up high CP detections. Circular polarisation at the level 3\% was found in the blazar 1519$-$293 \citep{Mac2000} and claimed to originate in a very compact source component. In a detailed high resolution study of the radio galaxy 3C84  \citet{HW2004} find similarly high levels of \pC\ in several jet components. It therefore appears, as was already claimed by \citet{Mac2000}, that high resolution is key for detecting high \pC. 

However, the large beams of our single dish survey certainly do not resolve the
AGN under study. High resolution observations at short mm wavelengths often find that the jet consists of the core and several knots \citep{Lee2008}, potentially making our study  vulnerable to beam--depolarisation.
Our finding of high  |\pC|$_{max}$ comparable to the values in high--resolution studies demonstrates that beam--depolarisation is in fact negligible in our sample, in agreement with the discussion in Sect.~\ref{ss:rates}.
We conclude then that the statistical trend of \pC\ increasing with frequency
is not primarily due to increasing resolution, but reflects an intrisic property of the CP generating mechnaism. We resume discussion of this 
finding in Sect.~\ref{ss:origin}.

\subsection{Fast CP variability }
\label{ss:variability}

Statistically, a source of our sample is detected at 3mm
7.2 times above $3\sigma$ during the 83 months of our monitoring program.
The mean time interval between detections is therefore nearly one year.
The {\it typical} source is thus detected at 3mm only once in 8 observations
during the brief period characterized by our sampling interval.
Given that the detection threshold of circular polarisation does not systematically vary during the monitoring period, the {\it fastest} time scale of CP variation is then of the order of our sampling interval of 
$\lesssim1$ month during the denser sampled period after 2009 (see also 
paper~I). 

Individual sources may occasionally depart from the mean picture.  There is however only one case in the whole database where S/N$\geq5$ detections
were made contiguously in time. This occured during 2007--2008 in 0836+710. In 0528+134, the source with the most detections, all of them are well separated by periods where \pC\ is below the detection threshold.
Inspection of  the \pC\ time sequences (Fig.~\ref{f:time1}) shows that this  is the rule. Noteworthy exceptions are discussed in sections \ref{ss:signs} and \ref{ss:events},  less prominent exceptions occur in 1226+023 and elsewhere.

The generally rapid variability  of CP is very different from the behavior of total flux density or linear polarisation degree and angle. We studied the relation between these
quantities using the discrete correlation function formalism which is described in detail in 
Paper~III. No significant (signal--to--noise ratio >3)
correlation was found  between CP and any of the other three  quantities.
These latter 3 quantities vary significantly more smoothly with time (Fig.~\ref{f:DBexample}
and 
Paper~III), demonstrating that our
time sampling is nearly adequate in Stokes $I$, and it is  mostly adequate for linear polarisation during the most prominent flares. However,
time variations of CP are clearly not resolved in our typical $\lesssim1$ month sampling interval.
A more precise estimate of the variation time scale requires dedicated and maybe more sensitive observations of a few selected sources.

Strong variability of CP was already noted in an early investigation by 
\citet{komesaroff1984} where they monitored all Stokes parameters of 22 AGN 
 with the Parkes telescope at 5 GHz during 6 years. 
These authors find that circular polarisation has the highest fractional variability of the Stokes parameters, and time scales are often shorter than 90 days, their average sampling interval. They explain these findings by  Faraday conversion driven by rotation of the {\bf B}--field direction
(BR--driven FC, see Sect.~\ref{ss:origin}). Their mechanism
 acts on a fraction of the cross section of a turbulent jet which may vary rapidly with time in a way that is uncorrelated with flux density or linear polarisation.   
This scenario is close to the modern turbulent extreme multi-zone (TEMZ) model
\citep{marscherTurbulence} in which the jet consists of a large number of randomly magnetized cells.  Their magnetic field can be partially aligned in a shock and circular polarisation can be generated through the BR--driven FC mechanism. \citet{MacDonald2016} have calculated the  transfer of polarised radiation in this situation and demonstrate that weak circular polarisation can indeed be generated at 43 GHz. 

Alternatively, the rapid CP variations may originate very close to the black hole in the fluctuations of the accretion flow like those shown in simulations of the jet/wind/disk system around a black hole \citep{McKinney2012}. The time scale for strong variations of the accretion rate in these simulations is found to be $\sim100\frac{r_g}{c}$, with $r_g$ being the gravitational radius and $c$ the speed of light. For a 
maximally rotating black hole of $10^9$ solar masses this is 5 days, and longer for slower rotation rates and lower masses. 
The current models of the jet/wind/disk system cover the immediate surroundings  of the black hole out to $\sim50r_g$. This is much smaller than the distance to the millimeter core of the jet which is thought to be located at 1-10 pc ($10^4 - 10^5 r_g$) from the black hole
\citep{Marscher2008,marscher1510,agudoOJ287,frommCTA102}. 
If the modelled variations can propagate from the disk all the way out to the millimeter core, they might induce the fluctuations of the orientation of the field needed for the rapid fluctuations of CP that we observe.

\subsection{Preference of handedness}
\label{ss:signs}

\begin{table}
\caption{Sources with strong preference for the sign of \pC\ at 3mm wavelength.
  The number of observations with positive/negative  sign 
  (right--hand--circular RHC / left--hand--circular LHC)
  is given for   detections at signal--to--noise ratios of 3 and 2. 
  Only sources with at least 10 detections are listed.}
\label{t:signs}      
\centering  
\begin{tabular}{l r@{\ }c@{\ }l r@{\ }c@{\,}l l} \hline
source   & \multicolumn{3}{c}{S/N $\ge3$} & \multicolumn{3}{c}{S/N $\ge2$} 
         & previous detections                    \vpfeiler{3.0ex}\\
 &  \multicolumn{3}{c}{+ / $-$} & \multicolumn{3}{c}{+ / $-$} &  \\[0.5ex]\hline    
0316+413 & \multicolumn{3}{c}{---} & 0 & / & 13 &[2,3,6] see text
                                                    \vpfeiler{3ex}\\
0355+508 & \multicolumn{3}{c}{---} & 1 & / & 11 \\ 
0528+134 & 15 & / & 1              &27 & / &  2 & RHC [2]\\ 
1127$-$145 &  0 & / & 10             & 1 & / & 20 & LHC [1]\\
1633+382 &  0 & / & 10             & 1 & / & 18 & LHC [5]\\ 
1641+399 & \multicolumn{3}{c}{---} &18 & / &  0 & RHC [4]\\
2251+158 & \multicolumn{3}{c}{---} &18 & / &  0 & RHC [1]; [4] see text \\[0.5ex]\hline
\end{tabular} 
\begin{tablenotes}
\item References:  (1)~\citet{komesaroff1984}; (2)~\citet{HW99}; (3)~\citet{aller2003};
   (4)~\citet{gabuzda2008}; (5)~\citet{Vitrishchak2008};
   (6)~\citet{HW2004} 
\end{tablenotes}
\end{table}

In the same investigation quoted above \citet{komesaroff1984} also
noted that despite the strong fractional variability of CP, reversals of the 
{\it  sign} of \pC\ were rare.
An even longer term sign consistency of \pC\ was observed
for the Galactic Centre black hole SgrA* (
\citet{bowerCpSgrA1999,saultMacquart1999,bowerCpSgrA2002,munoz2012};
Agudo et al, in prep.) where
circular polarisation was consistently found negative
(left handed).
Similar sign consistency was then reported for more AGN. \citet{homan2001}
found sign consistency for 5 out of 6 AGN,
 and attributed this remarkable property to  stability of the polarity
 of net magnetic flux in the jets. 

Motivated by these observations we looked at the stability of the  sign of
\pC\ in our sample of AGN. We find that at 3mm \pC\ had
positive and negative detections in the majority of sources.
Considering only sources with at least 10 detections (S/N $\ge3$) we find 29 such sources in our sample of 37. Only three have  a
clear preference for one sign (Tab.~\ref{t:signs}).  These ``quasi--unipolar'' sources have at most one detection of the discrepant sign. 
The source 0528+134 stands out with 16 detections, all but one  of
positive sign. 

We add in Tab.~\ref{t:signs} the detections made at the statistically less significant signal--to--noise ratio of 2. The 3 unipolar sources found above remain strongly asymmetric with the same sign preferred. In addition, four more sources appear to be strongly unipolar, including the quasar 2251+158
with 18 detections of positive sign and no negative ones. We consider these 7 sources as strong cases of unipolarity, but they constitute a 24 \% minority
of the 29 source with at least 10 detections.. 
Among these 7 sources, preference for positive or negative signs is roughly equally strong.       
The unipolar sources share this property with  our full sample
(Fig.~\ref{f:masterHisto}) where positive and negative signs of \pC\ are almost equally frequent. 

From our list of unipolar sources, four have been monitored previously \citep{komesaroff1984,HW99}.  The polarity of three of them (0528+134, 1127$-$145, 2251+158) is found in agreement, while for 0316+413 
the behaviour of CP appears to be more complex. A crossing of zero occured around 1980 \citep{aller2003} and \citet{HW2004} made the remarkable discovery of the simultaneous presence of jet components of opposite polarity. During our monitoring period, all but one observations with S/N$\geq1$ had negative CP (Fig.~\ref{f:time1}).   
Another type of complexity may occur in 2251+158 where CP was detected at 22 and 43 GHz by \citet{Vitrishchak2008}, albeit with different signs. 
Their sign at 43 GHz is positive, in agreement with our 18 observations at 86 GHz. It is conceivable that this change of sign, if real,  may be caused in this particular case by the jet opacity changing from optically thick to thin near 30 GHz if CP is generated by the intrinsic mechanism at this event.

Such opacity--induced sign flips must however be rare or even absent at our short wavelengths. The majority of our observations were conducted  with the 3 and 1mm bands in parallel (see 
Paper~I), permitting a reliable measurement of the spectral index $\alpha$ between these wavelengths at the time of the CP measurement. As shown in 
Paper~III, $\alpha$ is negative (mean of the sample is $-0.6$), indicating optically thin continuum emission.  Only in a few sources $\alpha$ reaches zero or small positive values for brief moments associated with total flux density peaks. We therefore discard opacity as a cause of CP sign flips.

In the jet model presented by \citet{gabuzda2008} the sign of \pC\ is
determined by the pitch angle and the helicity of the jet
{\bf B} field. The sign of \pC\ may then be expected to stay constant
as long as the basic geometry of the field does not change, as might be the case at the ejection of a new component of magnetised plasma.  From our data
(Tab.~\ref{t:signs} and Sect.~\ref{ss:signs}) this appears to be the
case only for one quarter of the AGN over the 7 year monitoring period.

\subsection{Origin of CP}
\label{ss:origin} 

Two mechanisms have been invoked for the generation of circular
polarisation in the jets of AGN \citep{WH2003}.  Firstly, weak circular
polarisation of the order of the inverse Lorentz factor of the relativistic
electrons is an inherent component  of  synchrotron radiation. Second,
CP can arise in the transport of the linearly polarised synchrotron
radiation through a non--uniform magnetic field (Faraday conversion).
Recent observations at wavelengths of 2cm or longer
\citep{wardleNature1998,Homan2009,osullivan2013} and theoretical
investigations \citep{rusz2002,beckertFalcke2002}  favor  Faraday
conversion as the dominant process.   

The degree of synchrotron--generated (intrinsic) CP at our wavelength of 3mm can be coarsely estimated from eq.~9 in \citet{WH2003} which considers a homogeneous jet with a partially ordered field. For a typical source in our sample we derive CP of the order of 1\%.   
This value is however likely an upper limit, because field reversals 
and a possible  positron jet component both decrease CP.
This suggests that synchrotron radiation is unlikely to generate the amount of CP observed {\it in the bulk} of our sample. We do not exclude, though, that the intrinsic mechanism is active or even dominant in cases with a high level of field order and optimum orientation.       

Our study therefore supports the generally held view that it is Faraday conversion which causes CP.
Faraday conversion in its simplest variant where, in homogeneous jets, conversion is driven by Faraday rotation (FR--driven FC) has 
however a very steep frequency dependence of $\nu^{-5}$ 
\citep{WH2003}, and thus appears to be ruled out by our findings.
However, following the discussion by these authors, BR--driven
Faraday conversion where the {\bf B} field changes direction along the jet may
have a flat or even inverted CP spectrum. This mechanism may be
expected to operate naturally in helical fields where it can be extremely efficient. Since jets of this field 
structure are believed to be common \citep{Gabuzda2004,Asada2008,Marscher2008,Gomez2008,2016ApJ...817...96G} 
Faraday conversion induced in helical fields appears to explain 
our finding of widespread and strong CP. 

 We caution that although we support BR--driven FC for the production of the CP observed in our sample, these previously discussed models are based on the assumption of homogeneous sources. For the more realistic case of inhomogeneous jets, like the  self--similar jet model discussed by \citet{BK1979},
the frequency dependence of CP may be very different. CP from the intrinsic mechanism as well as from Faraday conversion may then have flat or even inverted spectrum
\citep[see the discussion by][]{WH2003}.
Our findings of widespread (sect.~\ref{ss:rates}) and high CP (sect.~\ref{ss:weak}) at short mm wavelengths support such models where CP is not strongly frequency dependent. Furthermore, it may be also possible that the intrinsic mechanism as well as BR-- and FR--driven Faraday conversion operate simultaneously with relative contributions that change with time, be source dependent, or even change at different jet locations within the same source. In their detailed study of 3C\,279 using high resolution images at several radio frequencies, \citet{Homan2009} propose that some of the jet components have an inhomogeneous structure, and they tentatively attribute the CP detected in these components to the intrinsic mechanism. Unfortunately, our CP data at 1 mm are not yet accurate enough for deriving $m_{\text C}$ spectral indexes. Quasi-–simultaneous CP spectra covering radio and mm wavelengths are needed for more definite conclusions.

\subsection{Relation between linear and circular polarisation}
\label{ss:correlation} 

\begin{figure}
\centering
\includegraphics[width=0.48\textwidth]{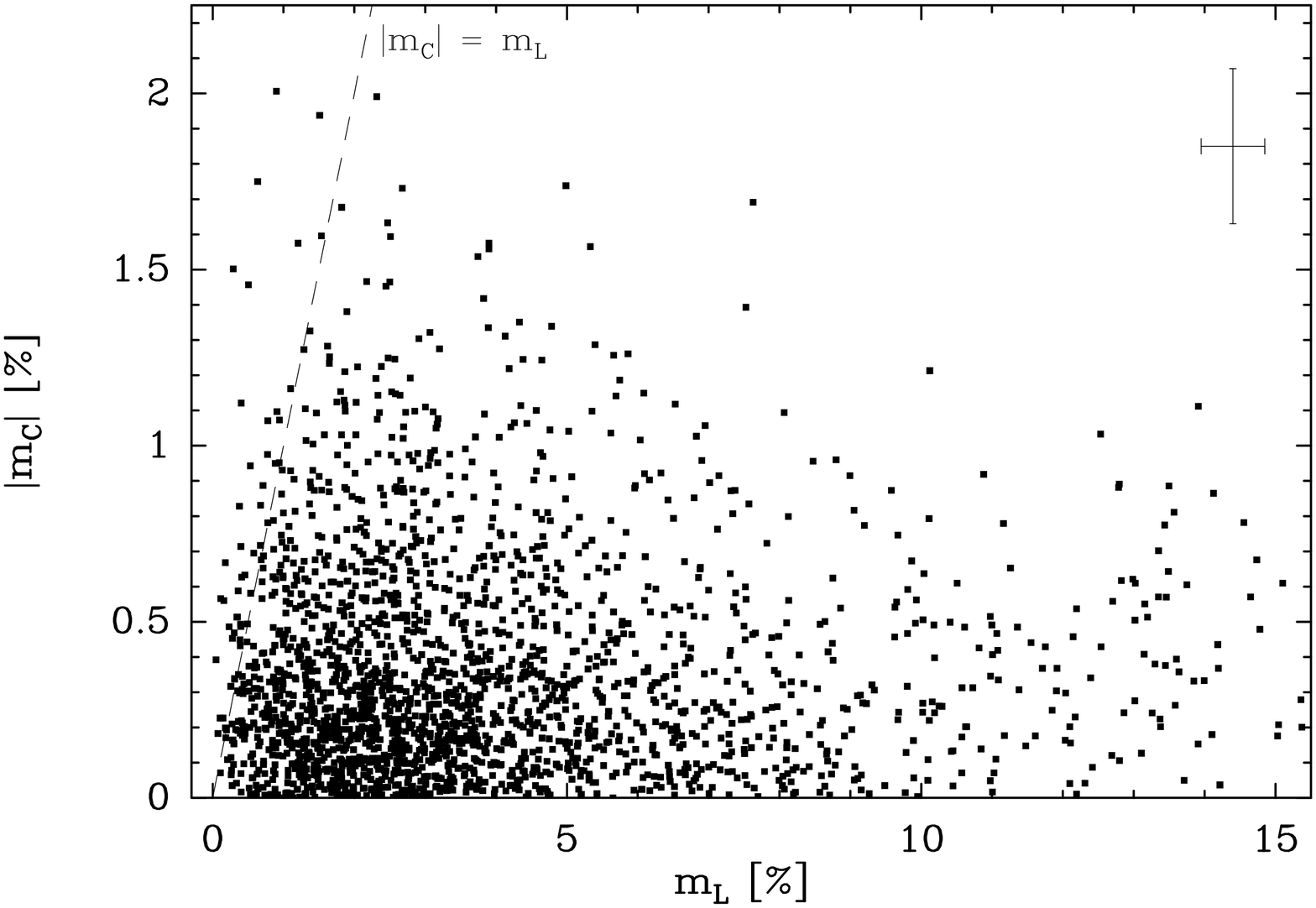}
   \caption{Circular polarisation, \pC, versus linear polarisation,
     \pL, at 3mm wavelength. The absolute value of \pC\ is plotted
     of all observations of all sources in our sample 
     (Tab.~\ref{t:bigTable}) where \pL > 0. 
     The cross bar at the right upper
     corner shows the typical observational error. 
     The dashed line indicates the locus where circular and linear
     polarisation are numerically equal. 
           }
      \label{f:tadpole}
\end{figure}

As we discussed in Sect.~\ref{ss:origin} that circular polarisation
may result from Faraday conversion of linear polarisation in regions of
changing magnetic field, it is interesting to explore the
statistical relation between \pC\ and \pL. Fig.~\ref{f:tadpole} shows
this relation for all 3mm observations of all sources in our sample. 
Since we did not see any statistical
difference between positive and negative \pC\ data, we
simply plot its absolute value |\pC| against the degree of linear
polarisation \pL. 

The observations are contained between the maximum |\pC|\ $\sim 2.0$\% 
as discussed above and the maximum \pL$\sim16$\%.
\footnote{Linear polarisation is discussed in 
Paper~III of this series
\citep{paperIII}}
Apart from a general trend toward low linear and circular polarisation, there
is no correlation between the two kinds of polarisation.  High
circular polarisation occurs at any value of \pL. The weak decrease
of the maximum |\pC|  toward high \pL\ may just be due to
a selection effect: there are fewer highly linearly polarised sources.  
The lack of correlation holds down to very low \pL\ values. Circular
polarisation as high as 1.5\% is detected in some sources where \pL\ is
below the $3\sigma$ detection threshold of 0.75\%.  \\

In Fig.~\ref{f:tadpole} we have drawn a dashed line where linear and circular
polarisations are equally strong. Half the linearly polarised flux is then converted into circular polarisation. Such high conversion efficiency is of course unlikely in AGN jets, given the randomness of magnetic field orientations, variable optical depths along the jet, and possible beam depolarisation. 
Furthermore, depolarisation due to an internal or external Faraday screen may also play a role at our frequencies if its rotation measure is high and variable \citep[see e.g. ][]{Attridge2005}.
Nevertheless, space near this dashed line
appears to be well populated by observations up to $\sim2.0$\%.   

There are 102 observations (4.5 \%) located to the left of the limiting line, most of which are not shown due to their clustering at \pL = 0.
In only 4 of these observations, however, CP is both significantly detected and significantly larger than \pL. Two of these observations were made on the quasar 0528+134.

Few other sources with this property have been found before: SgrA*
\citep{bowerCpSgrA1999,saultMacquart1999} and two low luminosity AGN,
M81* and NGC6500 \citep{bowerLLAGN2002}. A common feature of these
peculiar sources is the weakness or even absence of any jet. 
The 4 observations of AGN reported here 
do not easily fit into this picture. The explanation of the
 \pL<|\pC| property may  involve nearly complete depolarisation and 
 a substantial component of low energy electrons or thermal plasma near the mm core where efficient Faraday conversion can take place. In the model presented by \citet{Marscher2008}, the short millimetre emission originates in the turbulent zone of the jet upstream of the standing conical shock. Conditions favouring  \pL<|\pC| may then occur during periods when no strong shocks are propagating through this zone. Shocks would tend to align the magnetic field and thus likely increase linear polarisation. In sources with powerful jets, like those in our sample, such periods may be rare, while they may be the rule in more quiescent sources like SgrA*. 

\citet{rusz2002} and \citet{beckertFalcke2002} have theoretically studied these peculiar jets. In their models, the jet plasma is optically thick and highly turbulent. Complete depolarisation is avoided if the magnetic flux has a small bias and if there are not too many field reversals along the line of sight. The small bias of the magnetic flux also explains  the long term persistence of the handedness of CP.

\subsection{Events in CP time sequences}
\label{ss:events}

The dominant characteristic of the time sequences shown in Fig.~\ref{f:time1}  is their stochastic nature with only few observations above the 5$\sigma$ threshold. Comparing the typical 1$\sigma$ 
error bars with the distribution of the \pC\ values, it is also evident  that there are many more observations above 1$\sigma$ than expected from a Gaussian distribution. These characteristics have already been quantified in the histograms of the \pC\ distributions (Sect.~\ref{ss:histograms}), and they demonstrate the widespread presence of circular polarisation. But there are several events in the \pC\ time sequences which cannot be caught by a mere 
statistical analysis. We briefly describe two of them in this section, but postpone more detailed  analyses to later papers.
  
\subsubsection{A dip in 1127$-$145}
\label{sss:1127}
A nearly one year long dip of \pC\ occurs in this quasar centred on 2013.1. 
The dip is about 0.8\% deep. It is outlined by 12 observations, 6 of which are above $3\sigma$ and one being a $5\sigma$ detection. Most of the other observations outside the dip scatter around zero.

\begin{figure}
\centering
\includegraphics[width=0.48\textwidth]{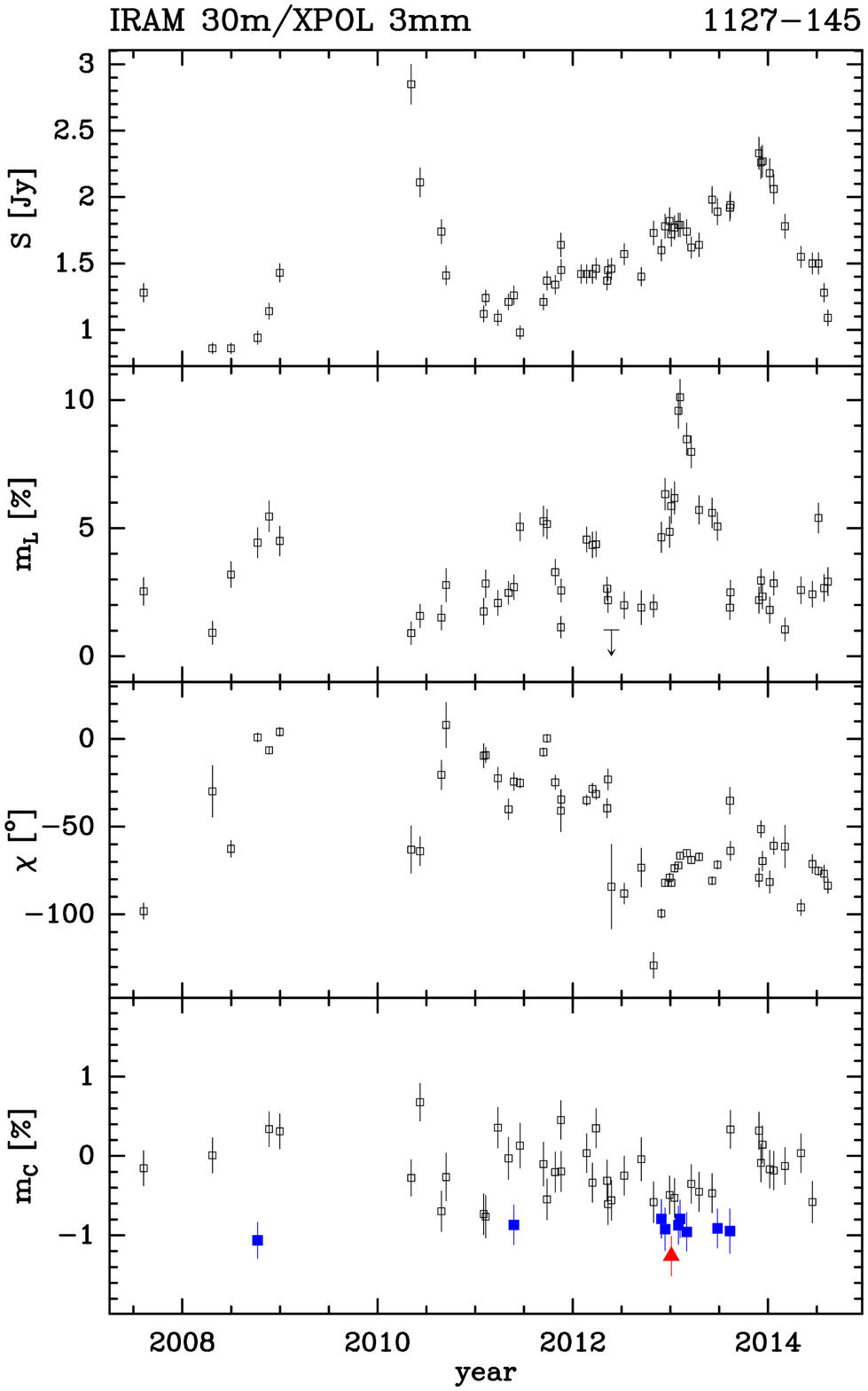}
   \caption{The quasar 1127$-$145 (z = 1.184). Detections of circular
            polarisation are shown as 
            filled symbols (blue squares for S/N$\geq3$ or red triangles
            for S/N$\geq5$). Data are from the POLAMI
            database.
           }
      \label{f:1127}
\end{figure}

Fig.~\ref{f:1127} shows that simultaneously to the \pC\ dip occurs a sharp peak in linear polarisation of similar  time duration and also peaking around the same time. A weighted mean of the angle measurement during the \pL\ peak
is $-80^\circ$, close to the position angle of the inner jet \citep{Jorstad2001, Tingay2002}. 
 The sudden and simultaneous  increase of linear and circular polarisation has no counterpart in the total flux light curve which is increasing almost featurelessly over 4 years between emission peaks near 2010.0 and 2014.0.  
There is no second observation of such simultaneous linear/circular polarisation surges in our database, nor are we aware of any such observation elsewhere. Such events are sufficiently rare that they get masked in \pL\ versus \pC\ plots like Fig.~\ref{f:tadpole} by the much larger number of situations when 
\pL\ and \pC\ are unrelated or \pC\ is simply below the detection threshold.

Such correlation is however a natural consequence of Faraday conversion, the favoured mechanism for generating circular polarisation. 
The conversion efficiency from linear to circular polarisation is of the order of 10\% during this  rather singular event. 
We speculate that the event may be caused by the emergence of a new jet component with a high degree of field order, but we postpone a more detailed discussion to a separate paper.

\subsubsection{Oscillatory behavior in 1055+018}
\begin{figure}
\centering
\includegraphics[width=0.48\textwidth]{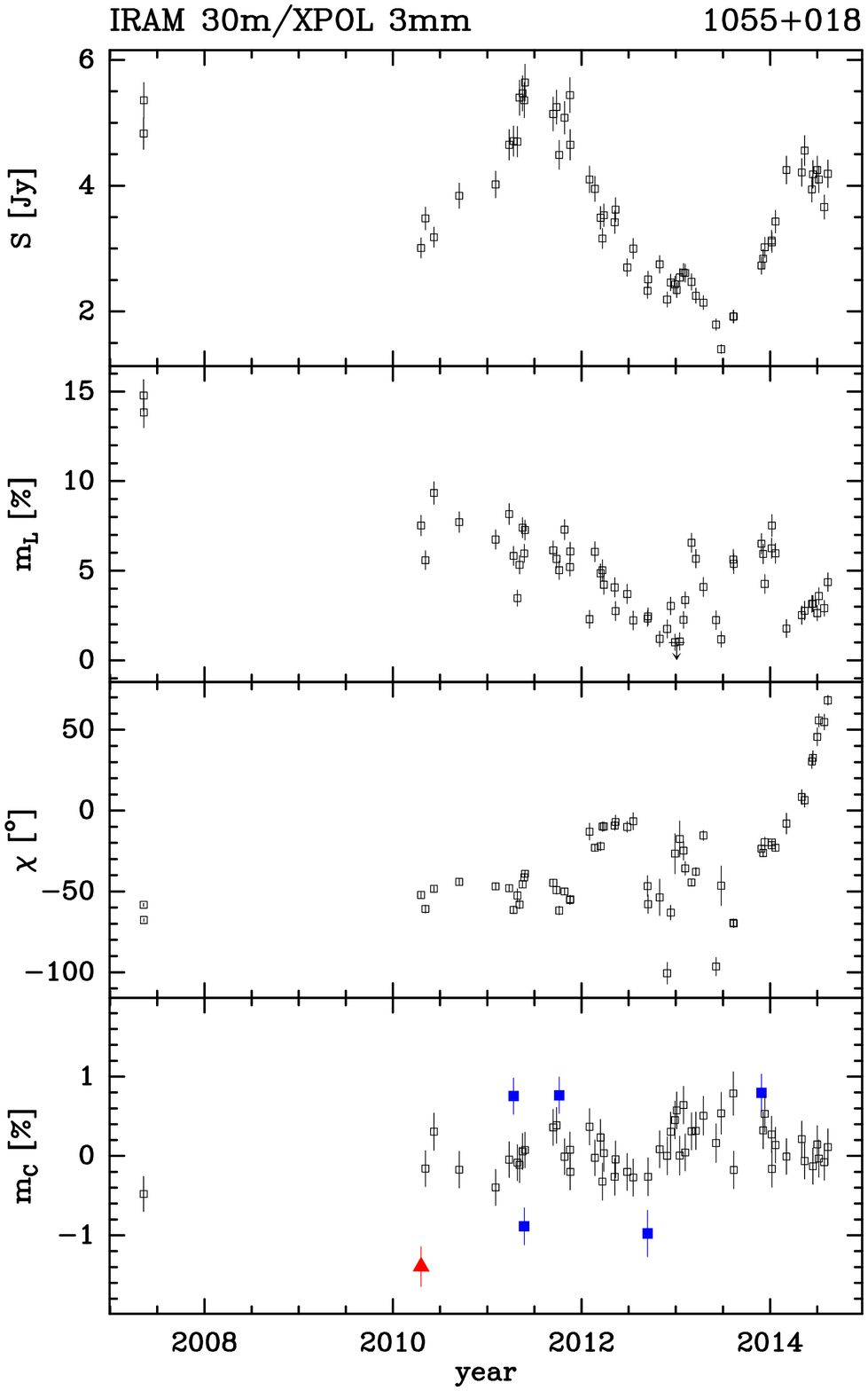}
   \caption{The BL\,Lac object $1055+018$ (z = 0.888).  
            Detections of circular polarisation are shown as 
            filled symbols (blue squares for S/N$\geq3$ or red triangles
            for S/N$\geq5$). Data are from the POLAMI database.
           }
      \label{f:1055}
\end{figure}

Systematic departures from \pC\ = 0 occur in several sources in our sample, but are most noticeable in the BL\,Lac object 1055+018 over a large fraction of the monitoring period. Departures go in positive and negative directions, and are reminiscent of oscillations. The intervals of systematic offsets from zero last $\lesssim\!1$ year, similar to 1127$-$145 discussed in the previous subsection. And as in the latter source, the excursions of \pC\ occur in the interval between 2 peaks of total flux density
where the emission evolves almost linearly without any significant  features. 
But unlike the event in 1127$-$145, the \pC\ excursions in 1055+018 do not have clearcut counterparts in linear polarisation (Fig.~\ref{f:1055}).

\subsection{Other CP properties of our sample}
\label{ss:other}

No correlations 
have been found between \pC\ and various
other source parameters. Specifically, there is no dependence of
\pC\ on AGN class (Fig.~\ref{f:masterHisto}) or redshift (not shown).
A systematic difference between quasars and BL\,Lac objects could have
followed from the difference in the angle under which the jets in both
classes   are viewed. The absence of such an effect is however
understandable from the argument that circular polarisation is due
to Faraday conversion which depends mainly on the structure of the magnetic
field in the jet. The very weak dependence on viewing angle is discussed in
\citet{gabuzda2008}. 

Furthermore, \pC\ does not systematically correlate with the simultaneously measured total flux density. 
The absence of any systematic correlation with linear polarisation was demonstrated in Sect.~\ref{ss:correlation}.

Despite the considerable number of \pC\ detections, we do not have a single observation where \pC\ was detected at the $5\sigma$ level in both the 3mm and 1mm bands. Higher sensitivity will be required for a reliable measurement of  the spectral slope of CP at short millimetre wavelengths.


\section{Conclusions}
\label{s:conclusion}

We present the results of our 7--year monitoring of circular polarisation (CP) of a sample of 37 AGN, the first such campaign at short millimetre wavelengths. The data, collected in the POLAMI database together with simultaneous measurements of linear polarisation and total flux density, demonstrate that CP is widespread in the blazar population and  easily detectable at 3mm. 
CP changes sign in most sources, but there is a subset of 7 sources whose CP polarity is strongly biased. 

The relatively easy detectability of CP at short mm wavelengths
constrains the mechanism by which AGN generate circular polarisation. The values at which we detect CP are comparable to the highest values observed at cm wavelengths. The intrinsic CP of synchrotron radiation appears to be too weak to account for the bulk of our CP detections. The same is true for CP generated by FR--driven Faraday conversion due to the strong frequency dependence ($\propto \nu^{-5}$) of this mechanism. We therefore favor   
BR--driven Faraday conversion where the linear polarisation of the 
synchrotron emission in one region of the jet can be  converted  to circular polarisation in a downstream region with different orientation of its {\bf B} field \citep{WH2003}.
The assumed widespread presence of helical magnetic fields in AGN jets 
combined with our finding of widespread CP in the blazars sample (all but one source detected at 3mm)
may provide a natural environment for this mechanism. The potentially high efficiency of CP generation in helical fields was already pointed out by these authors. Quasi--simultaneous CP spectra, ideally observed at high angular resolution and covering radio and mm wavelengths, are however needed for more definite conclusions.

A singular event observed in the quasar 1127$-$145 where around the year 2013.1 the quasar went through a nearly 1--year long period of correlated linear and circular polarisation. We speculate that the event may be caused by the emergence of a new jet component with a high degree of field order, but postpone a detailed interpretation to a future paper. 

The easy detectability of CP furthermore demonstrates that the large beam used in our observations is not a severe handicap. Apparently, at our high observing frequency only the jet core and very few more jet components are visible, possibly due to reduced opacity at short mm wavelengths. This greatly diminishes  the chances of self--cancellation of CP from components of different CP polarity inside the beam. 

We find that CP varies  at short ($\lesssim1$ month) and longer ($\sim1$ year) time scales. The fast variations of CP may either be caused by a high level of turbulence developing downstream in the jet \citep{marscherTurbulence} or originate in instabilities of the disk/jet system near the black hole as found in recent
simulations \citep{McKinney2012}. A characteristic time scale for a  $10^9$ solar masses black hole rotating near maximum this is 5 days.
If such variations propagate from the disk to the pc--scale jet, they might induce the fluctuations of the orientation of the field needed for the rapid fluctuations of CP that we observe.

The long time scale variations however cannot be explained in this way. 
The subset of sources (24\% of our sample) which have a strong bias in their CP polarity like 3C\,84 pose a similar problem. These sources may require more quiescent conditions in their jets, possibly approaching the situation described by 
\citet{Ensslin} where the sense of circular polarisation is directly related
to the sense of rotation of the black hole.

We cannot presently exclude that there exists in fact a continuous range of time scales, due to limitations of the sampling density and sensitivity of our observations. A dedicated program using an order of magnitude denser sampling and higher instrumental sensitivity is needed for a few of the most promising sources.

\section*{Acknowledgements}
The authors acknowledge the anonymous referee for his/her constructive revision of this paper.
We are grateful to Daniel Homan (Denison University, USA) for valuable comments and for sharing unpublished data, and to Ioannis Myserlis (MPIfR, Germany) for a careful reading of the manuscript. We thank an anonymous referee for insightful and helpful comments.
This paper is based on observations carried out with the
IRAM 30m telescope. IRAM is supported by INSU/CNRS (France), MPG
(Germany) and IGN (Spain). 
C.T. thanks the GILDAS software group, notably Sebastien
Bardeau, for competent and rapid support. 
I.A. acknowledges support by a Ram\'on y Cajal grant of the Ministerio de Econom\'ia, Industria y Competitividad (MINECO) of Spain.
The research at the IAA-CSIC was supported in part by the MINECO through grants AYA2016-80889-P, AYA2013-40825-P, and AYA2010-14844, and by the regional government of Andaluc\'{i}a through grant P09-FQM-4784.
This research has made use of the NASA/IPAC Extragalactic Database (NED) which is operated by the Jet Propulsion Laboratory, California Institute of Technology, under contract with the National Aeronautics and Space Administration.
We have also made use of NASA's Astrophysics Data System Bibliographic Services. 

\appendix
\section{Determination of the sign of Stokes V}
\label{s:crab}

The Stokes parameters $I, Q, U,$ and  $V$ have units of flux density, but unlike $I$, the parameters giving the polarisation $Q, U,$ and  $V$ can have positive and negative signs. The calibrations of their sign with XPOL has been briefly described in \citet{XPOL} for the previous generation receiver on the IRAM 30m telescope. Here we give a detailed description for the currrent receiver EMIR.

The most practical of the 3 methods described by \citet{XPOL} uses observations of the Crab nebula which is easily detected at 3mm.  Although the circular polarisation of the Crab nebula is negligibly small
at any radio wavelength (\citet{crabCP} and references therein), its linear polarisation at short millimetre wavelengths is high, \pL $\sim 30$ \%, and its angle is well measured, $\chi= 152^\circ$ \citep{crabLP}. Using a quarterwave plate (QWP) optimized for 86 GHz the linear polarisation is then readily converted to a strong circularly polarised signal. As soon as the relative orientation of the polarisation of source, receiver, and the optical axis of the plate are known, the distribution of the incoming linearly polarised flux into outgoing linearly and circularly polarised flux components and their orientation and sign is determined. 

\begin{figure}
\centering
\includegraphics[width=0.48\textwidth]{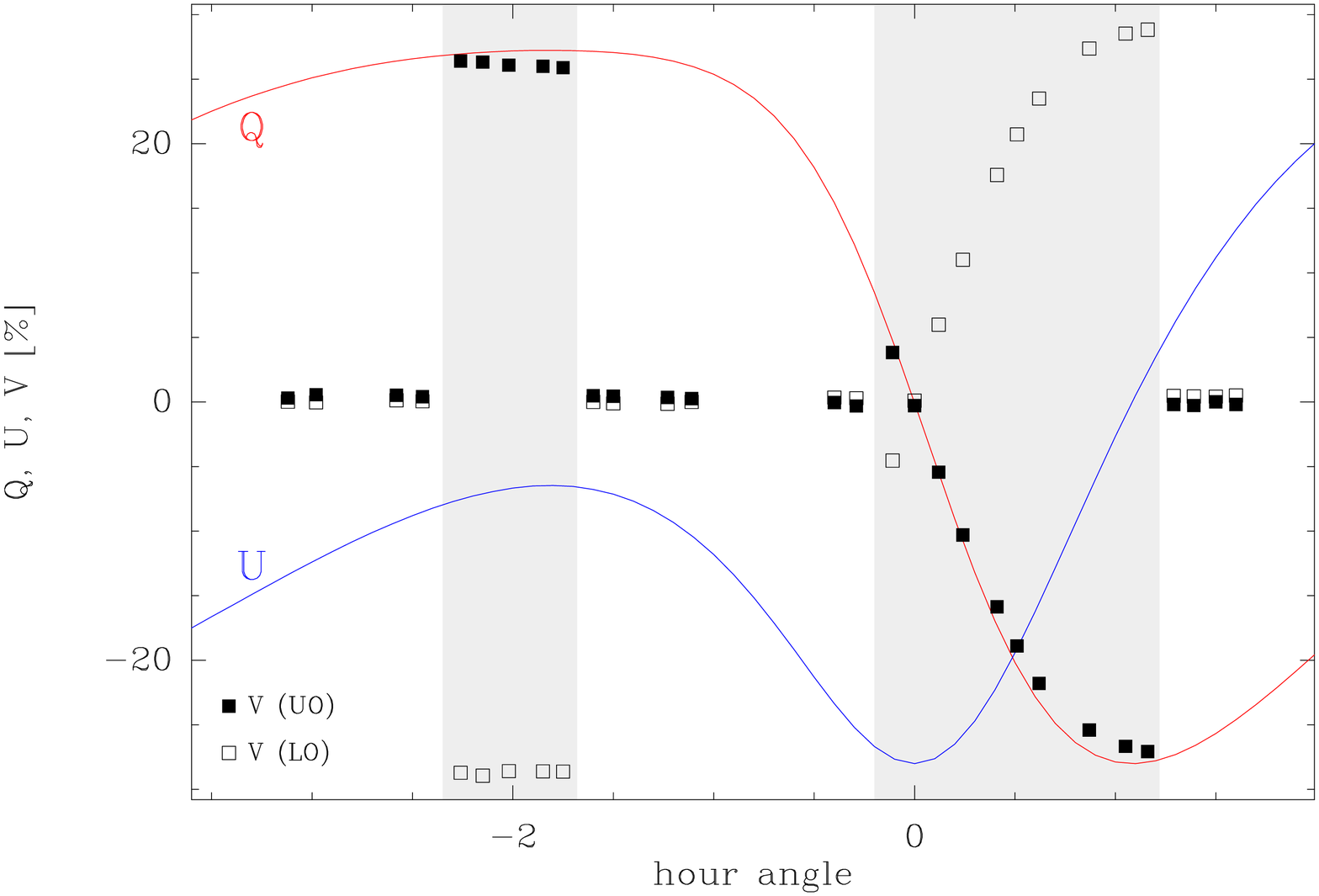}
   \caption{Circular polarisation of the Crab nebula as measured in the
            Nasmyth coordinate system $K_N$ versus hour angle. 
            Observations in both sidebands are shown. The grey shaded areas
            indicate periods when a quarter wave plate was inserted in the
            beam path. The coloured curves labelled $Q$ and $U$ indicate
            how the Crab nebula's linearly polarised flux projects into $K_N$.
           }
   \label{f:crab}
\end{figure}

Fig.~\ref{f:crab} shows observations of the Crab nebula in  Stokes $Q$ and $U$ over an hour angle range where $Q$ attains a broad maximum near HA = $-1.7$ and a sharper minimum near +1.1. Stokes $Q$ and $U$ curves refer to the Nasmyth coordinate system where the receivers are stationary and where the $x$--coordinate is horizontal and $y$ vertical. At HA =  $-1.7$ all linearly polarised flux is then in the horizontal component, and in the vertical component at HA = +1.2. 

We inserted a quarter-wave plate into the beam-path during the two periods which cover the two critical hour angles. The plate was oriented such that the
horizontal component H makes an angle of $45^\circ$ counted counterclockwise from the grooves as viewed from the receiver. In this configuration, incoming H is nearly completely converted into  lefthand circular polarisation (LHC) 
\footnote{We follow the IAU convention on polarisation which designates a wave as right hand circular (RHC) if its electric vector rotates counterclockwise when looking into the approaching wave. This convention also defines Stokes V = RHC -- LHC.
}. 
Stokes~V is thus at a negative maximum (LHC) at HA = $-2.0$ and at a positive maximum (RHC) at HA = +1.2 where the incoming linear flux is fully in the local vertical component. Since there is an even number (6) of reflections between the plate and the aperture plane, LHC at the plate is also LHC incident onto the telescope. Therefore, the QWP makes appear the Crab nebula  as a strongly LHC polarised celestial source when observed near HA = $-2.0$ and as RHC near HA = $+1.2$.

These strongly circularly polarised waves are seen by the linearly polarised channels of EMIR as waves whose vertical component is advanced/retarded by $90^\circ$ at the critical hour angles HA = $-$2.0/+1.2. Since XPOL derives Stokes~V from cross-correlation of the two channels after downconversion into one of its four IF sub-bands \citep{EMIR}, we must consider how downconversion affects the relative phase between the two wave components.

Frequency downconversions maintain the phase relationship of the original waves at radio frequency if downconversion occurs into the upper sideband (USB), but the relative phase is inverted when downconverted into the lower 
sideband (LSB).  
Table~\ref{t:IF} summarises the phase relationship resulting for the 4 EMIR sub-bands. A positive sign indicates the phase between two incoming waves at radio frequency is maintained after downconversion, and inverted for the sub-bands with negative signs. 

We have made observations of the Crab nebula at 86 GHz using EMIR configured to record the lower-outer (LO) and upper--outer (UO) sidebands simultaneously.  XPOL observations were made during hour angles from $-3.0$ to $+1.5$. At or near the critical hour angles the QWP was located in the beam as described above. The results obtained are shown in Fig.~\ref{f:crab} where the periods with QWP inserted are greyed. 

Outside the greyed periods the observed Stokes~V is zero in both sidebands as it should be \citep{crabCP}.
Inside the first greyed area near hour angle $-2.0$ apparent circular polarisation is as strong as the linear polarisation observed at 86 GHz
\citep{crabLP}. The Crab's linearly polarised flux is nearly fully converted into circular as expected. We find Stokes~V negative for the LO subband and
positive for the UO subband. In the second greyed window near HA= +1.2 these signs are reversed, and as discussed above. Since we expect LHC (Stokes~V negative) near HA = $-2.0$, the subband LO gives the correct result. According to Table~\ref{t:IF}, the signs of the subbands LI and UO must be inverted, while the sign of UI is preserved. The EMIR Stokes~V data which we discuss in this paper were all obtained in the LI subband, and their sign has consequently been inverted in accordance with  Table~\ref{t:IF}.

\begin{table}
\caption{Phase relationship with respect to radio frequency for the EMIR subbands.
}
\label{t:IF}      
\centering  
\begin{tabular}{l c c c c}\hline
subband           & LO   & LI  & UI  & UO \\[0.6ex]\hline
IF frequency, GHz &$-10$ &$-6$ & +6  & +10 \\
phase relation    & +    & $-$ & $+$ & $-$ \\[1ex]\hline
\end{tabular}
\end{table}


\bsp	
\label{lastpage}
\end{document}